\documentclass[onecolumn]{emulateapj}
\submitted{Accepted for publication in the Astrophysical Journal June 26, 2009}
\shortauthors{RANGWALA \& WILLIAMS} \shorttitle{FABRY-PEROT SPECTROSCOPY OF GALACTIC BAR: METALLICITIES}
\begin{document}

\newcommand{\ca}         {\mbox{\rm \ion{Ca}{2}} $\lambda 8542$}
\newcommand{\feh}          {\textrm{Fe/H}}
\newcommand{\kms}          {km s$^{-1}$}
\newcommand{\cole}          {$W_{20}$}
\newcommand{\car}          {$W_{40}$}
\newcommand{\voi}         {$W_{\infty}$}
\newcommand{\sumca}        {$\sum W$}
\newcommand{\wpr}          {\mathrm{W}^{\prime}}

\title{Fabry-P\'{e}rot Absorption Line Spectroscopy of the Galactic
  Bar. II. Stellar Metallicities} \author{Naseem
  Rangwala\altaffilmark{1} and T. B. Williams\altaffilmark{1,2}}
\altaffiltext{1}{Department of Physics and Astronomy, Rutgers
  University, 136 Frelinghuysen road, Piscataway, NJ
  08854}\altaffiltext{2}{Visiting astronomer, Cerro Tololo
  Inter-American Observatory, National Optical Astronomy Observatory,
  which are operated by the Association of Universities for Research
  in Astronomy, under contract with the National Science Foundation.}
\begin{abstract}
We measure the \ca\ line strength in 3360 stars along three
lines-of-sight in the Galactic bar: $(l,b) \sim (\pm 5.0,
-3.5)$\degr\ and Baade's Window, using FP absorption line
spectroscopy. This is the first attempt to show that reliable
absorption line strengths can be measured using FP spectroscopy. The
\ca\ line is a good indicator of metallicity and its calibration to
          [Fe/H] is determined for globular cluster red giants in
          previous investigations. We derive such a calibration for
          the bulge giants and use it to infer metallicities for our
          full red clump sample (2488 stars) at all three
          lines-of-sight. We present the stellar metallicity
          distributions along the major axis of the bar. We find the
          mean $\mathrm{[Fe/H]} = -0.09 \pm 0.04$ dex in Baade's
          Window, and find the distribution in this field to agree
          well with previous works. We find gradients in the mean
          metallicity and its dispersion w.r.t Baade's WIndow of $-0.45$
          and $-0.20$ dex respectively at $l = +5.5$\degr\, and of $-0.10$
          dex and $-0.20$ dex at $l = -5.0$\degr. We detect a signature
          of a possible tidal stream at $l = +5.5$\degr, in both our
          velocity and metallicity distributions. Its radial velocity
          indicates that it is not associated with the Sagittarius
          stream. We also measure the metallicity of a bulge globular
          cluster NGC 6522 in our Baade's Window field to be $-0.90
          \pm 0.10$ dex, in agreement with recent measurements of
          \citet{zoc08}. This agreement demonstrates the reliability
          of our metallicity measurements.
\end{abstract}
\section{Introduction}
This is the second paper of the series on measurements of radial velocities and equivalent widths
using the \ca\ line for a large sample of stars in the Galactic bar, based on Fabry-P\'{e}rot (FP)
absorption line spectroscopy. In Paper I \citep{rangwala09}, we used FP imaging spectroscopy at
three lines-of-sight (LOS) in the bar: $(l,b) \sim (\pm 5, -3.5)$ and Baade's Window (BW) to
obtain a total of 2488 bar red clump giants (RCGs), 339 bar M giants (red giant branch), and 318
disk main sequence stars. This sample is an order of magnitude larger than any previous sample for
a given LOS.

The existence of a bar in the Milky Way (MW) is now firmly established. Its photometric signature
is seen as a distinct peanut-shaped bulge in the COBE maps \citep{dwek95,cobe98}. The signature of
the bar is also confirmed in the bulge red clump giant (RCG) population \citep{stanek97}, in
microlensing \citep{alcock00}, and in stellar and gas kinematics \citep{howard08, rangwala09,
weiner99}. These morphological and dynamical observations strongly indicate a bar-like feature,
formed from secular evolution processes. But detailed chemical abundance measurements
\citep{rich88,matteucci92,ballero07,zoccali07} in Baade's Window find an old, $\alpha$-enhanced,
and metal-rich population, indicating a bulge that was formed on a very short timescale with high
star formation efficiency. To understand and resolve the issues related to the formation of
structure in the inner Galaxy, a variety of dynamical and chemical studies have been performed
over many years and still continue. For recent progress in the studies of the Galactic bulge see
the reviews by \citet{minniti08} and \citet{rich0710}.

In Paper I, we showed that FP techniques are very efficient for simultaneously measuring high
precision radial velocities of a large sample of stars. Our large radial velocity sample produced
an accurate determination of the detailed shape of this velocity distribution in the bar and
measured the difference in the radial streaming motions of the stars on the near and far sides of
the bar. These observations provide strong constraints on the dynamical models of the bar
(Debattista et al. in preparation).

The main goal of this paper is to test the suitability of FP spectroscopy for reliable and robust
line strength measurements, and to determine the stellar metallicity distributions along the three
LOS. The calcium triplet (CaT) lines in the cool giants have been established as reliable
indicators of [Fe/H], and have been used extensively to infer the metallicities of globular
clusters \citep{az88, ad91, zinn85, rutledge97, cole04,carrera07} and of dwarf spheroidal galaxies
\citep{tolstoy05}. These investigations have determined the calibration between the summed
equivalent widths of the triplet and [Fe/H] using red giants in numerous globular clusters (GC)
over a range of [Fe/H] ($-2.0 \lesssim \mathrm{[Fe/H]} \lesssim +0.5$ dex). We can apply similar
techniques to use our \ca\ line strength measurements to obtain [Fe/H] for our large sample of
stars. We demonstrate the reliability of our EWs by measuring the mean [Fe/H] of NGC 6522, a bulge
globular cluster that is present in our BW field. We also determined a new calibration between the
CaT index and [Fe/H] for the bulge RCGs by using recent high resolution [Fe/H] measurements by
\citet{zoccali07}, and apply this calibration to infer metallicity for our sample.

Measuring the stellar metallicity distribution (SMD) from a large sample of stars can be a very
useful tool for the chemical evolution models to constrain different formation scenarios for the
bulge (e.g. Did it form inside-out or outside-in? What is the timescale of its formation?). The
SMD in BW has been measured by many authors using both high and
low-resolution spectroscopy and photometric data \citep{sadler96,fulbright06,zoccali03,
zoccali07}. These measurements in BW find the bulge to be metal-rich and $\alpha$-enhanced,
supporting the formation model where the bulge was formed on a short timescale (less that a Gyr)
with a high star-formation rate \citep{ballero07}. If there is a bar and a classical bulge then
abundance studies in BW alone will not be sufficient to separate the two components. Looking for
metallicity gradients along the bar can provide additional constraints on differences between the
two components. Our measurements differ from previous works both in the large sample size and in
including two positions along the major axis of the bar in addition to BW.

The combination of velocity and metallicity distributions can be used to detect the signature of
tidal streams associated with disrupted satellites. The Sagittarius stream was discovered in the
measurements of the bulge velocity distribution by \citet{ibata95}. We report in this paper an
indication of a possible tidal stream at our $l = +5.5$\degr\ LOS.
\section{Observations and Data Reduction}
The observations were made using the Rutgers Fabry-P\'{e}rot system on the CTIO 1.5m telescope. A
total of 10 sub-fields were observed: four each at $l \simeq \pm 5$\degr\ (MM7B and MM5B fields
respectively) and two in BW. The positions of these fields are listed in Table 1 of Paper I. We
chose to observe the \ca\ line, which is one of the strongest features in the late-type stars like
the RCGs, is not contaminated by the foreground ISM, and is relatively free of strong terrestrial
emission lines. We used a medium-resolution etalon with a spectral response function that is well
fit by a Voigt function with a FWHM of 4 \AA, equivalent to approximately 140 \kms\ at 8500 \AA.
We typically scanned a range of 8530 -- 8555 \AA\ with wavelength steps of 1 \AA.

The basic procedure in Fabry-P\'{e}rot (FP) imaging spectroscopy is to obtain a series of
narrow-band images, tuning the interferometer over a range of wavelengths covering a spectral
feature of interest. The data cube thus produced is analyzed to extract a short portion of the
spectrum of each object in the field of view, using DAOPHOT \citep{stetson87} to measure a flux of
each star in each image. For a field of view with many objects of interest, the technique can be
extremely efficient. For example, in this work we obtained about 500 stars per data cube for our
4\arcmin\ field of view.

\begin{figure}
\center
 \includegraphics[scale=0.65]{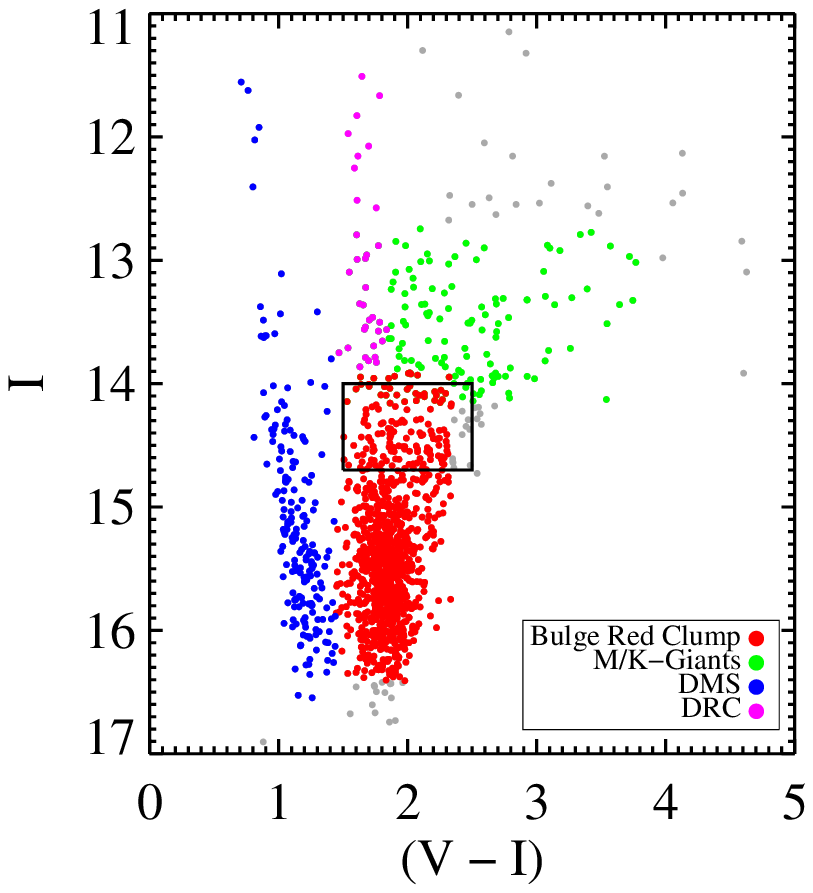}
  \includegraphics[scale=0.5]{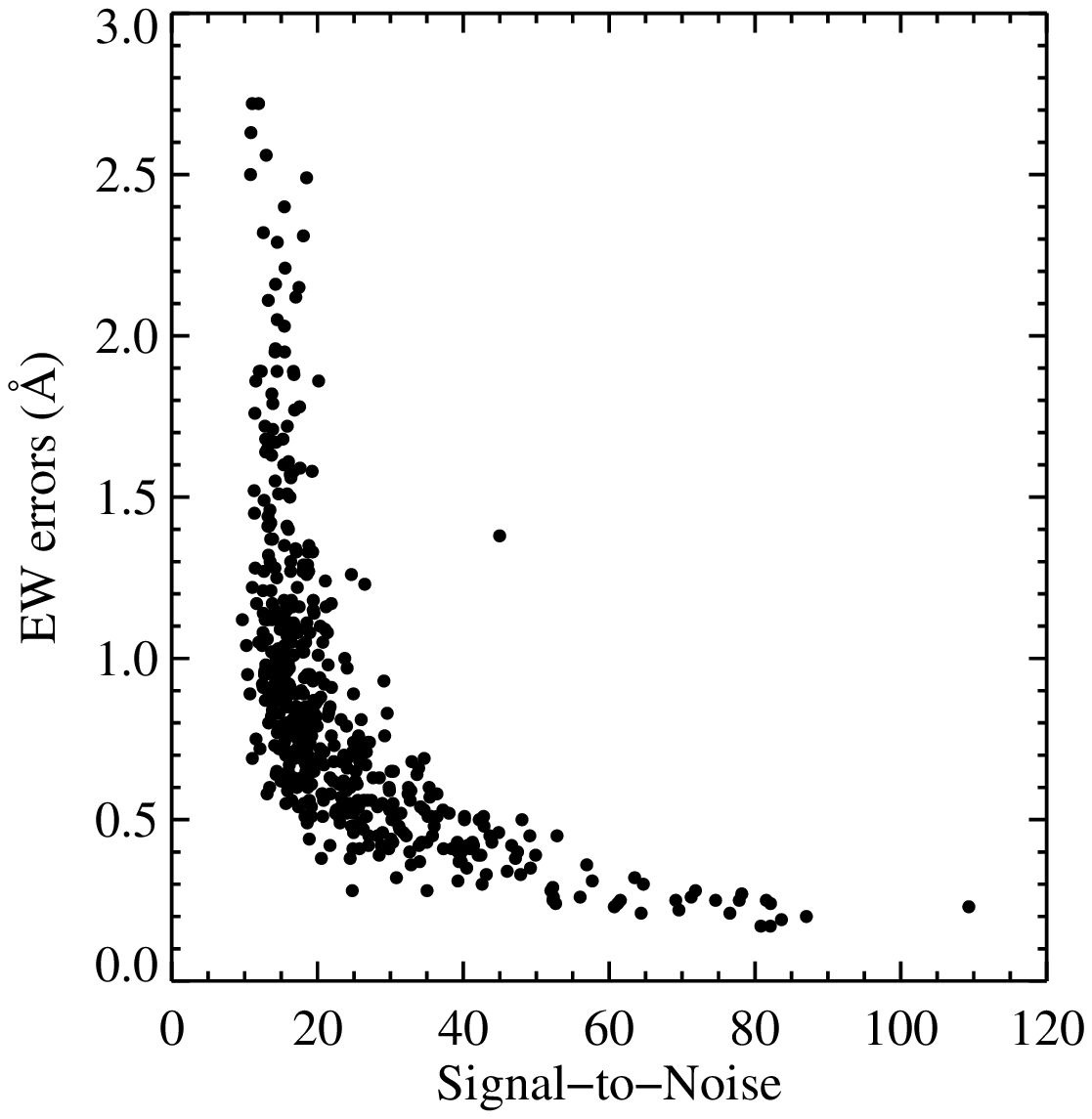}
\caption{Left: CMD for $(l,b) = (-5.0, -3.5)$ LOS. We have radial velocity and
$\mathrm{EW_{8542}}$ for every star on this plot. CMDs at the other two LOS have similar features.
The rectangle shows the bulge sample selection of \citet{zoccali07}. The I and V band photometry
comes from the OGLE survey. Right: EW uncertainties as a function of S/N for $l = -5$\degr\
line-of-sight.}\label{cm}
\end{figure}

\begin{figure}
  \includegraphics[scale=0.8]{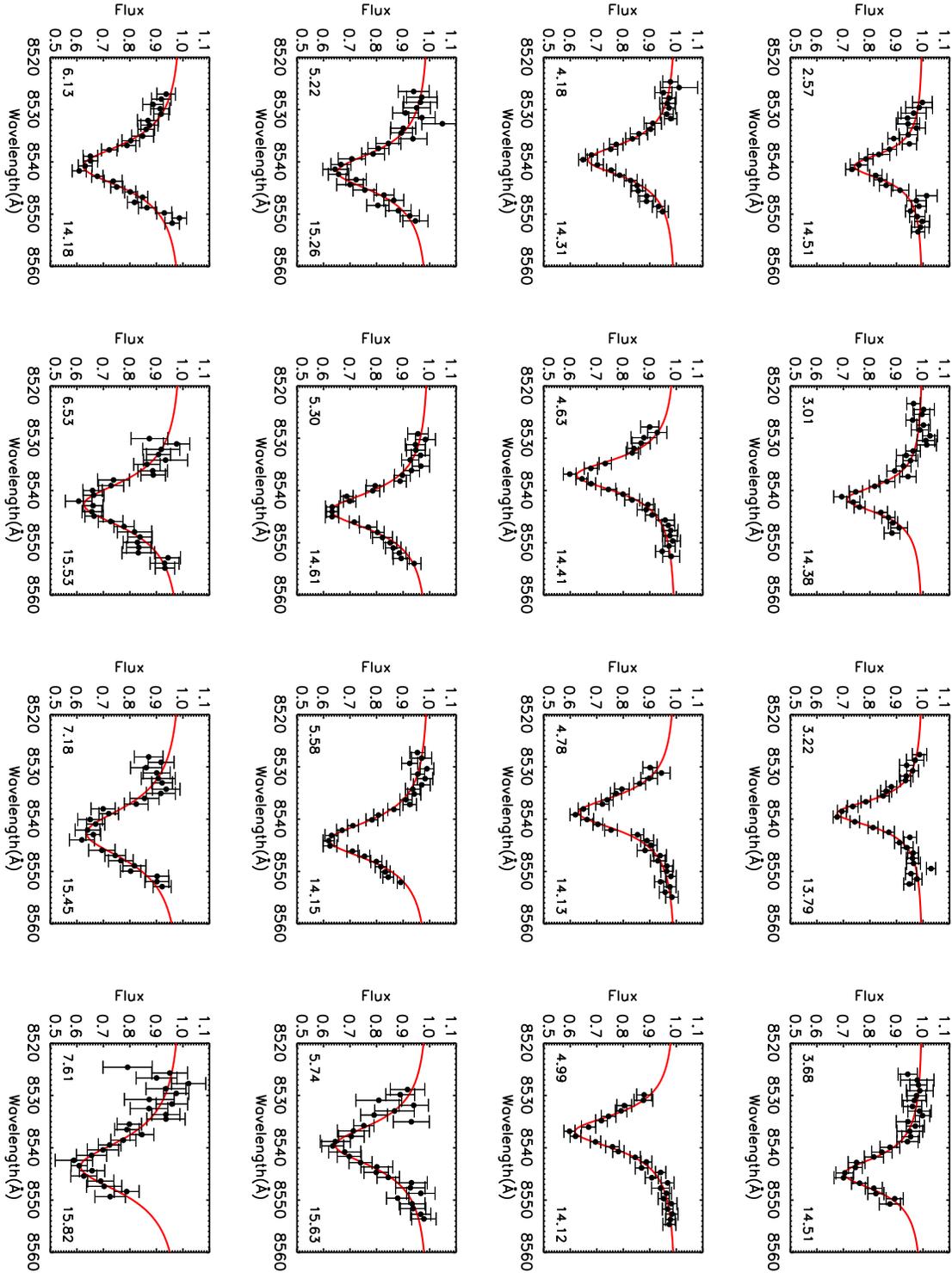}
  \caption{A sub-sample of \ca\ absorption lines illustrating the range of EWs measured. Solid line is the Voigt fit.
  EW (lower left) and I-band magnitude (lower right) are listed for each spectrum.}\label{spec}
\end{figure}

The \ca\ absorption line is fit with a Voigt function to obtain the
central wavelength, continuum, total area in the line, and the
Gaussian and Lorentzian widths. A sub-sample of our spectra are shown
in Figure \ref{spec}, illustrating the range of equivalent widths (EW)
measured. Our sample for three LOS consists a total of 2488 bar RCGs,
$\sim 300$ bar M-giants and $\sim 350$ disk main sequence stars. To
select these various stellar populations we used I and V band
photometry, the majority of which comes from the OGLE catalogue
\citep{database}. The selections are shown in Figure \ref{cm} (left
panel) and discussed in Paper I. As shown in \citet{cenarro01}, in
later-type M-giants a TiO band distorts the continuum around the
calcium triplet, and strongly affects the measured equivalent width of
the lines. Our color selection limits our sample to K-giants, and
avoids this effect.

\begin{figure}
    \center
  \includegraphics[scale=0.55]{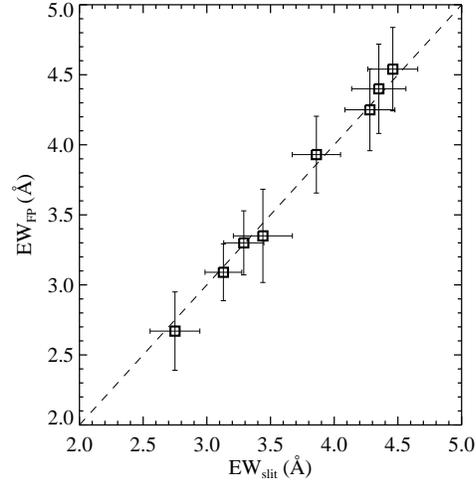}
  \caption{Comparing EWs of \ca\ line measured by fitting Voigt function to 25 \AA\ bandpass (FP) and 60 \AA\
  bandpass (slit).}\label{fpslit}
\end{figure}

The right panel of Figure \ref{cm} shows the uncertainties of the EW measurements over the S/N
range of our sample. The median fractional uncertainty is 1/6th of the EW. As expected, the
uncertainty increases when the S/N decreases, but there is a wide uncertainty range at S/N $\sim
20$, which comes mostly from sampling different portions of the line in different stars (see
Figure \ref{spec}). Another source of uncertainty comes from the fact that we observe a limited
part of the spectrum around the \ca\ line compared to slit spectra, which typically measure the entire
line profile and surrounding continuum. To investigate the effects of our limited spectral range,
we generate 1000 random realizations of the spectrum of each of 8 red giants from the globular
cluster 47 Tuc (provided by Andrew Cole; private comm.). We fit Voigt functions to these spectra
over two different bandpasses: 25 \AA\ (FP) and 60 \AA\ (slit). The mean EW and standard deviation
for each of the 8 stars is plotted in Figure \ref{fpslit}. No bias is introduced in the EW due to
the limited FP bandpass, but the uncertainty is larger by a factor of 1.5.

We only measure the single, strongest line of the triplet. The total EW of all three lines is
$\sum\mathrm{W} = \mathrm{W}_{8498} + \mathrm{W}_{8542} + \mathrm{W}_{8662} =
2.21\mathrm{W}_{8542}$. The factor of 2.21 is verified using the data from 47 Tuc and was also
confirmed by Andrew Cole (private comm.), who finds this factor to be the same over a wide range
of metallicities. This factor is also consistent with synthetic spectra, the solar spectrum
\citep{Mould76}, and the spectra of many dwarf and giant stars \citep{jones84}.
\section{Measurement of Equivalent Width and Comparison with Different Indices}
In Paper I, we published measurements of the EW of the \ca\ line by fitting a Voigt function to
the absorption line and integrating the fit over all wavelengths. We call this index \voi\ to
indicate that the integration bandpass is from 0 to $\infty$. Our spectra are measured over a
limited wavelength range, which could affect the continuum and EW measurements. In addition we
need to assess the effects of our \voi\ definition, which differs from the indices used in
previous investigations. We use both the synthetic spectral line database of
\citet{munari}\footnote{http://archives.pd.astro.it/2500-10500/} and also spectra of the red
giants from 47 Tuc to analyze these effects.

We compare our index to those of \citet{cole04} (C04 hereafter) and \citet{carrera07} (Ca07
hereafter). C04 measure the EW of all three of the calcium triplet lines, fitting each with the
sum of a Gaussian and a Lorentzian, and then integrating the fit over a 20 \AA\ bandpass. Ca07 use
the same functional form as C04 but different, and wider continuum and line
bandpasses\footnote{The line and continuum bandpass used by Ca07 are defined in
\citet{cenarro01}}. Ca07 found no significant differences between their index and that of C04, but
we will show below that there is a small systematic difference between them. These two indices
differ from \voi\ in three ways: the functional form, the continuum definition, and the
integration bandpass. We address the effects of each of these factors below.

We created a series of artificial profiles using a Voigt function with a range of Gaussian and
Lorentzian widths, and then fitted them with the sum function. In every case the sum fit is
indistinguishable from the Voigt profile and yields the same EW. We also verified that fitting the
47 Tuc spectra with the two functional forms gives identical profiles and EW. We conclude that the
value of the EW index is independent of the choice between these two functional forms. We choose
to use the Voigt function because the intrinsic stellar line profile is a Voigt, our instrumental
profile is a Voigt, and the convolution of the two is also a Voigt.

We generate synthetic spectra with the following stellar parameters: $T_{\textrm{eff}} = 4750$ K,
$\textrm{log (g)} = 3.0\, (\mathrm{in}\,\, \textrm{cm~s}^{-2})$, $\textrm{V}_{\textrm{rot}} = 5\,
\textrm{km\, s}^{-1}$, micro-turbulence = 2 \kms\ and $[\alpha/\textrm{Fe}] = 0.0$. These values
are consistent with \citet{zhao01}'s high-resolution spectroscopic measurements of the red clump
giants in the solar neighborhood. Synthetic spectra were obtained for [Fe/H] = [-2.0, -1.5, -1.0,
-0.5, 0.0, +0.5] dex, covering the entire range of metallicities available in the database. We
then convolve these synthetic spectra with our instrumental profile, a Voigt of $\sim 4$ \AA\
FWHM.

\begin{figure}
 \center
 \includegraphics[scale=0.65,angle=90]{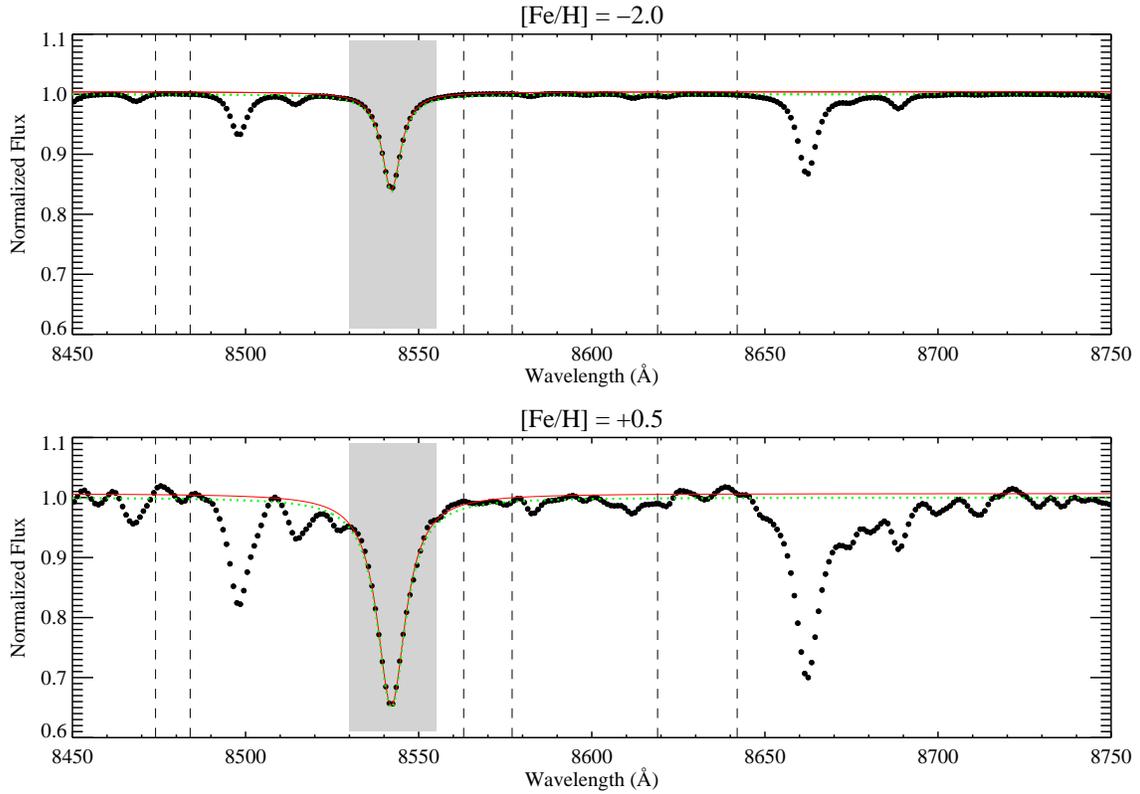}\\
  \caption{Synthetic spectra for two different metallicities are fit with a Voigt function using two different
  estimates of the continuum: (1) the continuum bands from Ca07 (red solid line) and (2) the continuum from the best-fit
  Voigt function (green dotted line). The shaded region shows the range (8530 \AA\ -- 8555 \AA) over which we fit the
  Voigt function to our FP spectra. Vertical dashed lines show Ca07 continuum bands. \label{ss}}
\end{figure}

We fit the \ca\ line in the synthetic spectra with a Voigt function in two ways: first, in which
the continuum is measured as the mean flux within several bands, using exactly the same procedure
and definition as Ca07, and second, in which the continuum is one of the parameters of the Voigt
function fit over a 25 \AA\ bandpass from 8530 -- 8555 \AA\ to simulate our FP spectra. We fitted
the spectra for all available metallicities (listed above), and show the fits for the highest and
lowest [Fe/H] in Figure \ref{ss}. The solid and dotted lines show the fits with the two different
continuum definitions. The continuum windows of Ca07 are shown by vertical dashed lines. The
differences in the two continuum levels are about 0.4\% and 0.7\% for the low and high [Fe/H],
respectively and the FP-fit continuum is always lower than the Ca07 continuum. For an integration
bandpass of 40 \AA\ the continuum difference would produce an EW difference of less than 0.3\AA\ in
even the strongest-lined stars. We will neglect this systematic difference, both because our
random errors are relatively large (median error $\sim 0.7$ \AA), and because our calibration will
compensate for systematic effects.

\begin{figure}
 \center
 \includegraphics[scale=0.6,angle=90]{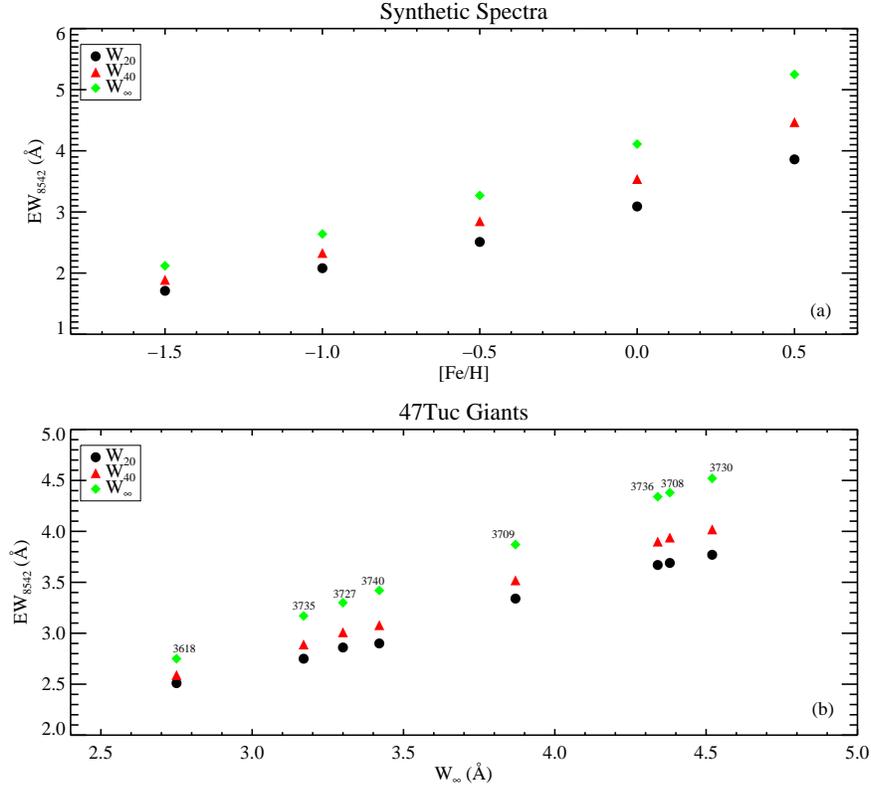}\\
  \caption{Top panel: \voi (diamonds), \car\ (triangles) and \cole\ (circles) are measured using synthetic spectra and plotted as a function of
  [Fe/H]. Bottom panel:  The three indices are plotted as a function of \voi\ using spectra of 8 red giant stars
   from 47 Tuc. The points are labeled with the star ID.}\label{47tuc}
\end{figure}

\begin{figure}
\centering
  \includegraphics[scale=0.4]{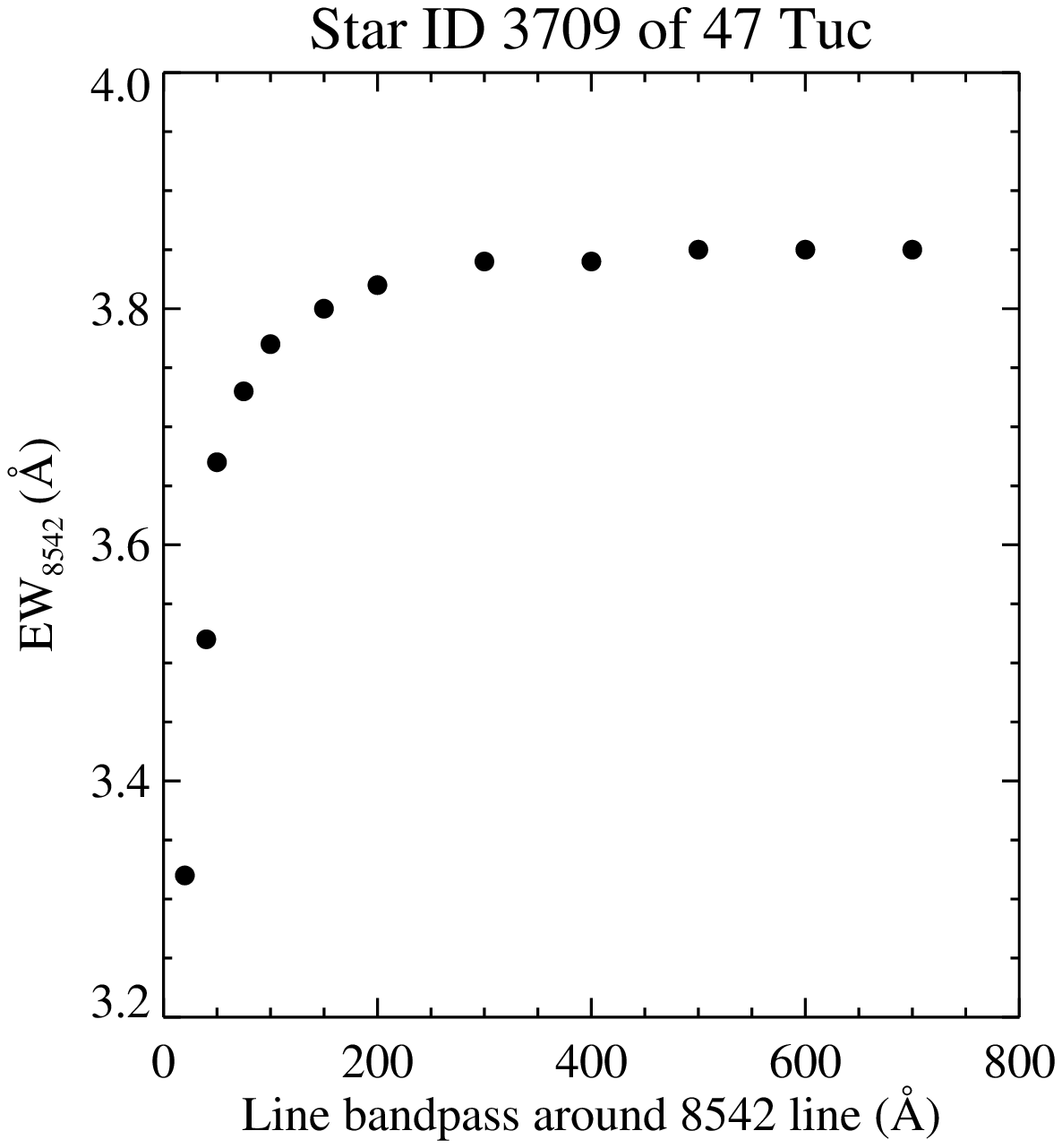}
  \includegraphics[scale=0.4]{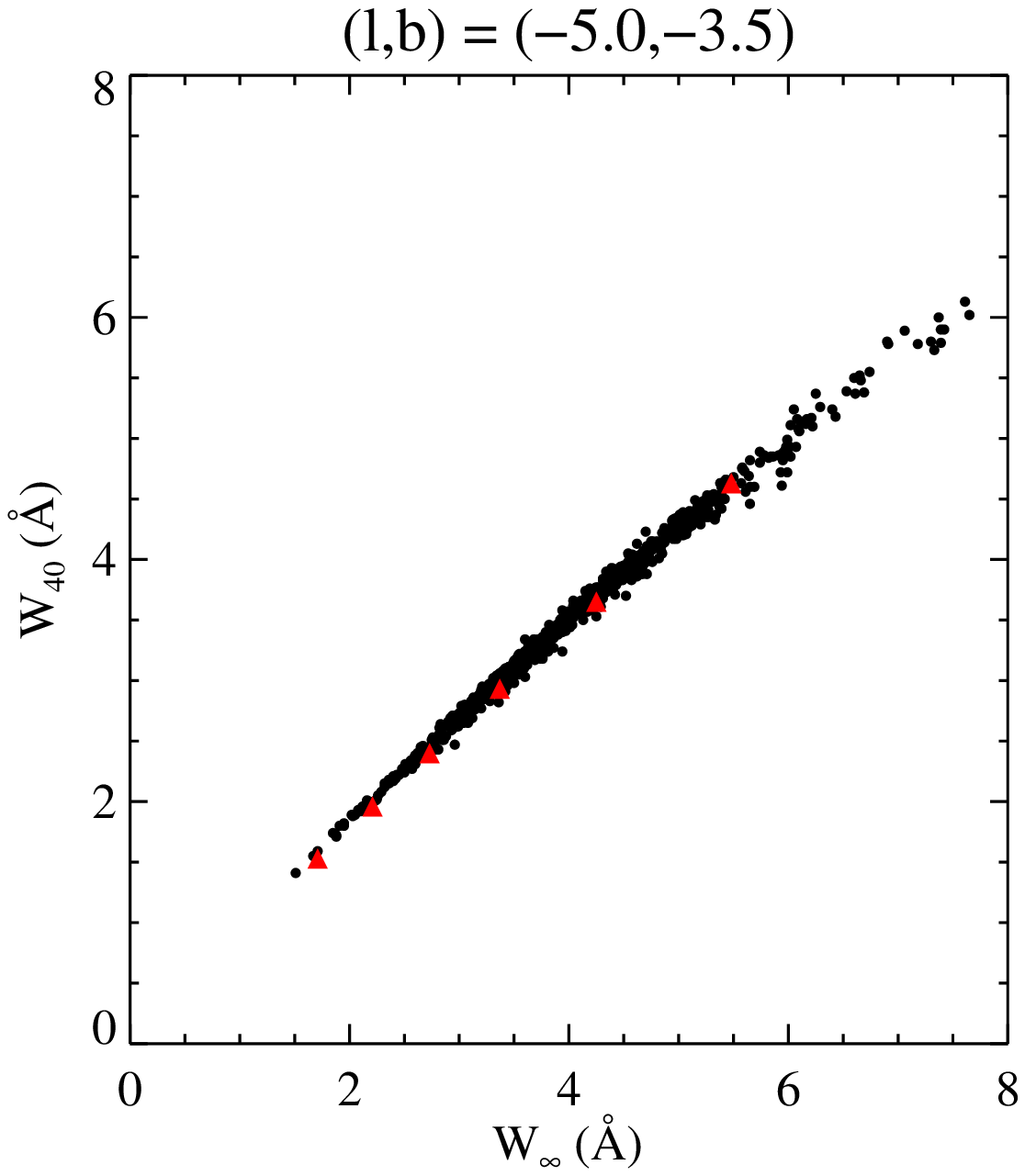}
  \includegraphics[scale=0.4]{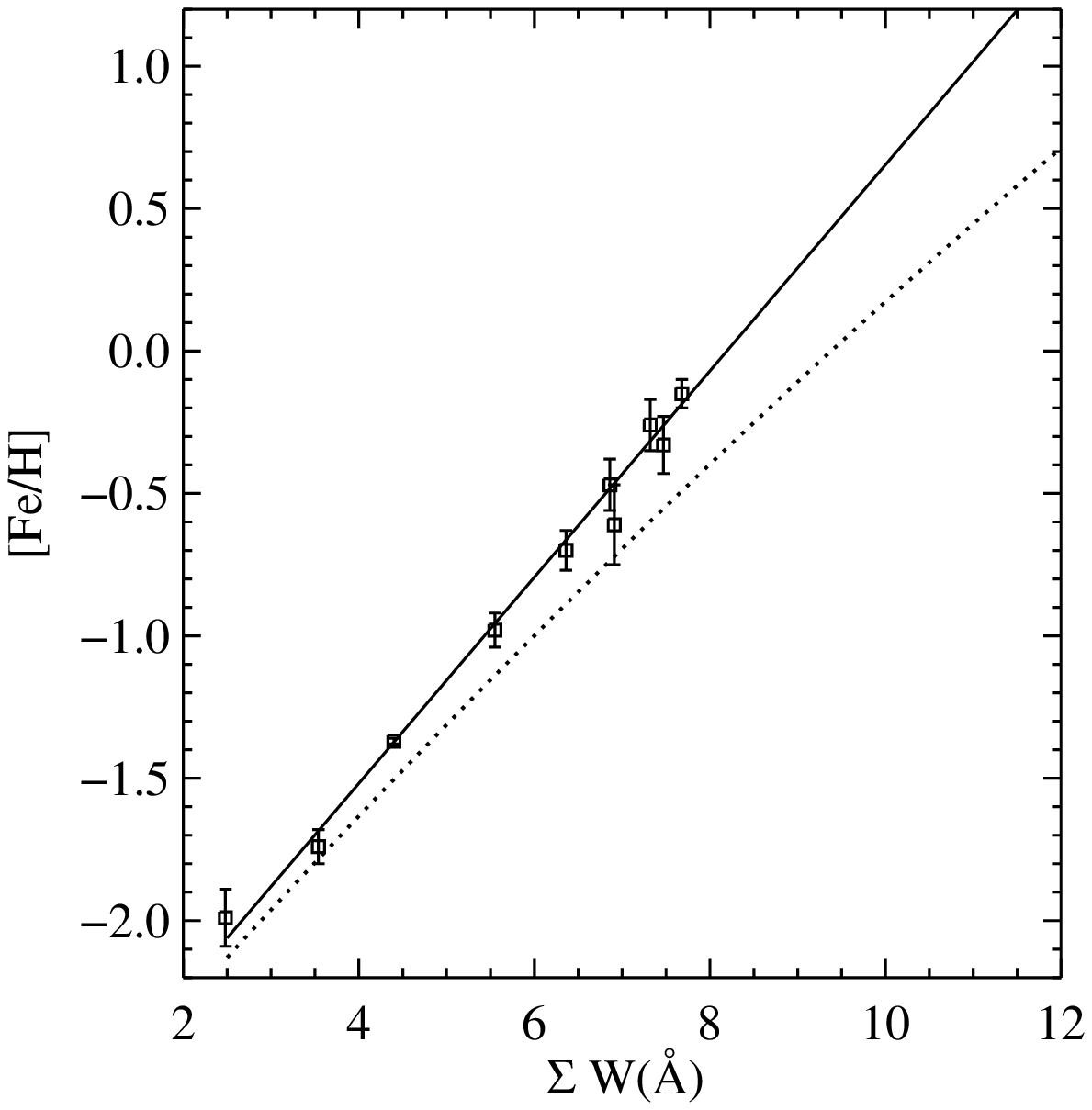}
  \caption{Left panel: EW as a function of line bandpass. Middle panel: \voi\ and \car\
  for our sample at $l = -5$. Red points are from synthetic spectra. Right Panel: Solid line: calibration of C04's index for a
  sample of globular clusters; dotted line: Calibration curve transformed to our index $W_{\infty}$.}
  \label{deviation}
\end{figure}

Since the choice of fitting function and the continuum definition have at most a minor effect on
the line index, the choice of integration bandpass is the major contributor to the systematic
differences between the index values. Collisional or pressure broadening in the stars' atmospheres
gives rise to strong damping wings on the calcium line absorption profiles. In metal rich stars,
as much as 75\% of the EW comes from the wings \citep{mendes89}, and in the Sun the wings extend
out to at least 30 \AA\ on either side of the line center \citep{smith88}. Thus we expect that the
measured values of the line indices will depend strongly on the spectral range over which the
fitted profiles are integrated.  We will denote the bandpass definitions of C04 and Ca07 by \cole\
and \car\, respectively, and our integration of the entire profile as \voi. Figure \ref{47tuc}
compares these indices for the \ca\ line, measured in synthetic spectra with a range of
metallicities, and also in the observed spectra of 8 red giants in 47 Tuc.  Broader integration
bandpasses produce systematically larger EW index values, and these differences increase with line
strength (or metallicity). In the left panel of Figure \ref{deviation} we show the measured
equivalent width of the \ca\ line in one of the 47 Tuc stars as a function of integration
bandpass.  The first two points in the plot show \cole\ and \car, respectively.  The EW converges
to \voi\ for bandpasses greater than 200 \AA.

In the middle panel of Figure \ref{deviation} we compare the values of \voi\ and \car, measured
for our entire sample of FP spectra at $l=-5$\degr. As expected from the tests shown in Figure \ref{47tuc},
there is a tight, but non-linear, relation between the indices. We overplot the indices from the
synthetic spectra of Figure \ref{47tuc}a, showing that the indices measured from the limited-range
FP data are in excellent agreement with those measured from more extensive, traditional spectra.
We have used Monte Carlo simulations to estimate the uncertainties of \voi\ and \car\ in the fits
to our FP data, and find that the fractional uncertainties of the two indices are identical. Thus
it seems to make little difference which integration bandpass one chooses to use. Of course, to
apply a previously-determined calibration of the CaT index to [Fe/H], one must use the same index
definition. In the right panel of Figure \ref{deviation} we show the calibration of \cole\ for
globular clusters as measured by C04. We use the \voi\ $-$ \cole\ correlation to calculate the
corresponding calibration relation for \voi, shown by the dotted curve.
\section{Comparison with High-Resolution Metallicities}
\subsection{Metallicity of NGC 6522}
NGC 6522 is a bulge globular cluster that appears at the southwest edge of our BW field (see
Figure 2 of Paper I). The stars from this cluster show up as a distinct kinematic feature in our
velocity distribution (see Figure 8 of Paper I). We can use the EWs of these stars to measure the
mean metallicity of this cluster via the calcium infrared triplet method. We will compare our
measurement to the most recent high-resolution [Fe/H] determination for this cluster by
\citet{zoc08} (Zoc08 hereafter), in order to assess the accuracy of our procedures.

\begin{figure}
\center
  \includegraphics[scale=0.65]{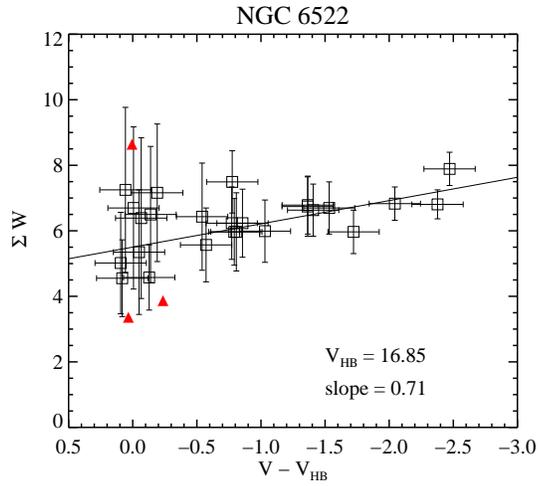}\\
  \caption{The summed EW varies linearly with V band magnitude. The solid line is the best linear fit. Red triangles
  indicate excluded stars.}\label{trend}
\end{figure}

To isolate possible members of this cluster from the background bulge population we use the
kinematics and photometry of the cluster. We identify the red giant branch of this cluster from
the HST CMD \citep{piotto02} and select the stars within 1-$\sigma$ (7.9 \kms) of the cluster's
mean velocity ($-14.67$ \kms). We find 28 possible members of this cluster, majority of which fall
within 2\arcmin\ of the cluster center. Compared to the mean EW at a given V magnitude, the EW of
three stars deviates significantly (one with a relatively large EW and two with relatively small
EWs). Because the cluster's velocity is very near the center of the bulge velocity distribution,
the kinematic selection constrain cannot eliminate all of the bulge contamination. In addition,
all three stars are near the faint end of the photometric distribution, where there is a higher
probability of bulge contamination. We therefore exclude these three stars from the cluster
sample. The three excluded stars are indicated by triangles in Figure \ref{trend}.

To convert EWs to metallicity we follow the method in C04. For this analysis we will use \car\ so
that we can consistently use the calibration of C04 and compare to previous work. The first step
is to measure the `reduced equivalent width' to correct for the strong effects of stellar
effective temperature and surface gravity in the red giants. This is given by:
\begin{equation}\label{rew}
\mathrm{W}^{\prime} = \sum\mathrm{W} + \beta(\mathrm{V} - \mathrm{V_{HB}})
\end{equation}
where the introduction of the horizontal branch magnitude ($V_{\mathrm{HB}}$) removes dependence
on the cluster distance or reddening, and $\sum\mathrm{W} = 2.21$\car\ is the summed EW of all
three CaT lines. Figure \ref{trend} shows this relationship for $V_{\mathrm{HB}} = 16.85 \pm 0.20$
\citep{rutledge97}. $\sum\mathrm{W}$ increases linearly up the red-giant branch as expected. The
best fit to the data is shown by the solid line with a slope of $\beta = 0.71 \pm 0.22$ \AA\
mag$^{-1}$.

Within the 1-$\sigma$ error our slope is in good agreement with previous measurements of
\citet{rutledge97}, Ca07 and C04 who measured it to be $0.64 \pm 0.02$, $0.647 \pm 0.005$ and
$0.73 \pm 0.04$ respectively, using the data from numerous globular clusters.

The calibration between $\mathrm{W}^{\prime}$ and [Fe/H] as measured by C04 is:
\begin{equation}\label{cal}
\mathrm{[Fe/H]} = (-2.966 \pm 0.032) + (0.362 \pm 0.014)\mathrm{W}^{\prime}
\end{equation}
This calibration has a rms scatter of $\sigma = 0.07$ dex and is measured for metallicity range of
$-2.0 \lesssim \mathrm{[Fe/H]} \lesssim +0.5$ dex \citep{carrera07}.

Using this calibration and $\mathrm{W}^{\prime} = 5.70 \pm 0.15$ \AA\ (for $\mathrm{V_{HB}} =
16.85$) we obtain $\mathrm{[Fe/H]}_{\mathrm{NGC6522}} = -0.90 \pm 0.10$ dex . For another value of
$\mathrm{V_{HB}} = 16.52 \pm 0.07$ as measured from the HST CMD of \citet{sosin97} for this
cluster we get $\mathrm{W}^{\prime} = 5.96 \pm 0.15$ \AA\ and $\mathrm{[Fe/H]} = -0.81 \pm 0.10$
dex.

We compare this result with Zoc08 where they measure high-resolution abundances for stars in the
Galactic bulge using VLT FLAMES-GIRAFFE spectrograph. Their sample consists of seven members from
NGC 6522. The mean [Fe/H] from these seven stars is $-0.91 \pm 0.06$ dex. Our measurement is in
excellent agreement with this work. There is one star (OGLEID 412752) in common with both Zoc08
and our sample. For this star we get [Fe/H] = $-0.84$ dex and $-0.94$ dex ($\pm 0.42$ dex) for
lower and higher values of $\mathrm{V_{HB}}$ respectively, in agreement with $-0.80 \pm 0.15$ dex
as measured by Zoc08.

We also compare our result to \citet{rutledge97}, who measured the
calcium triplet index for 72 GCs. They get a lower [Fe/H] of $-1.21
\pm 0.04$ using 9 possible members of this cluster. They use different
weights for the EWs of the three calcium lines to get $\wpr = 3.47 \pm
0.11$.  Converting our index to their system using Equation 3 of Ca07
we get $\wpr = 4.08 \pm 0.12$, which does not agree with Rutledge's
index. Our index converted to their system gives [Fe/H] $= -0.91$ dex
using their calibration between $\wpr$ and [Fe/H]. Because NGC 6522 is
a bulge cluster and has a velocity close to the mean of the bulge
velocity distribution, it is very difficult to establish the
membership for this cluster. From a sample of 11 stars that
\citet{rutledge97} select as cluster members, 8 stars show velocity
deviations larger than 20 \kms\ from the mean cluster velocity,
suggesting that their sample may be significantly contaminated by
bulge stars.
\subsection{\ca\ -- [Fe/H] calibration for the bulge stars}
\begin{figure}
\center
  \includegraphics[scale=0.5]{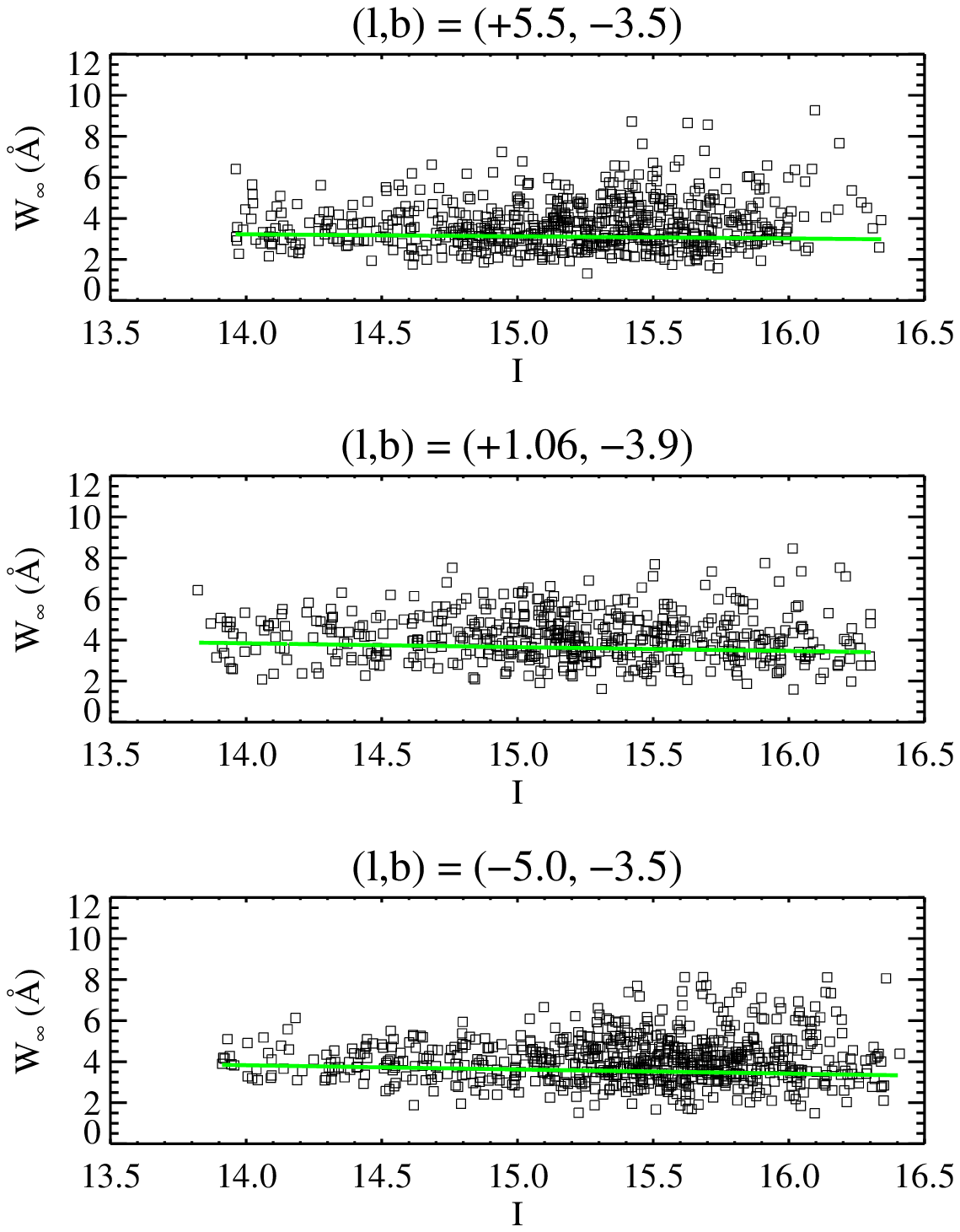}
  \includegraphics[scale=0.5]{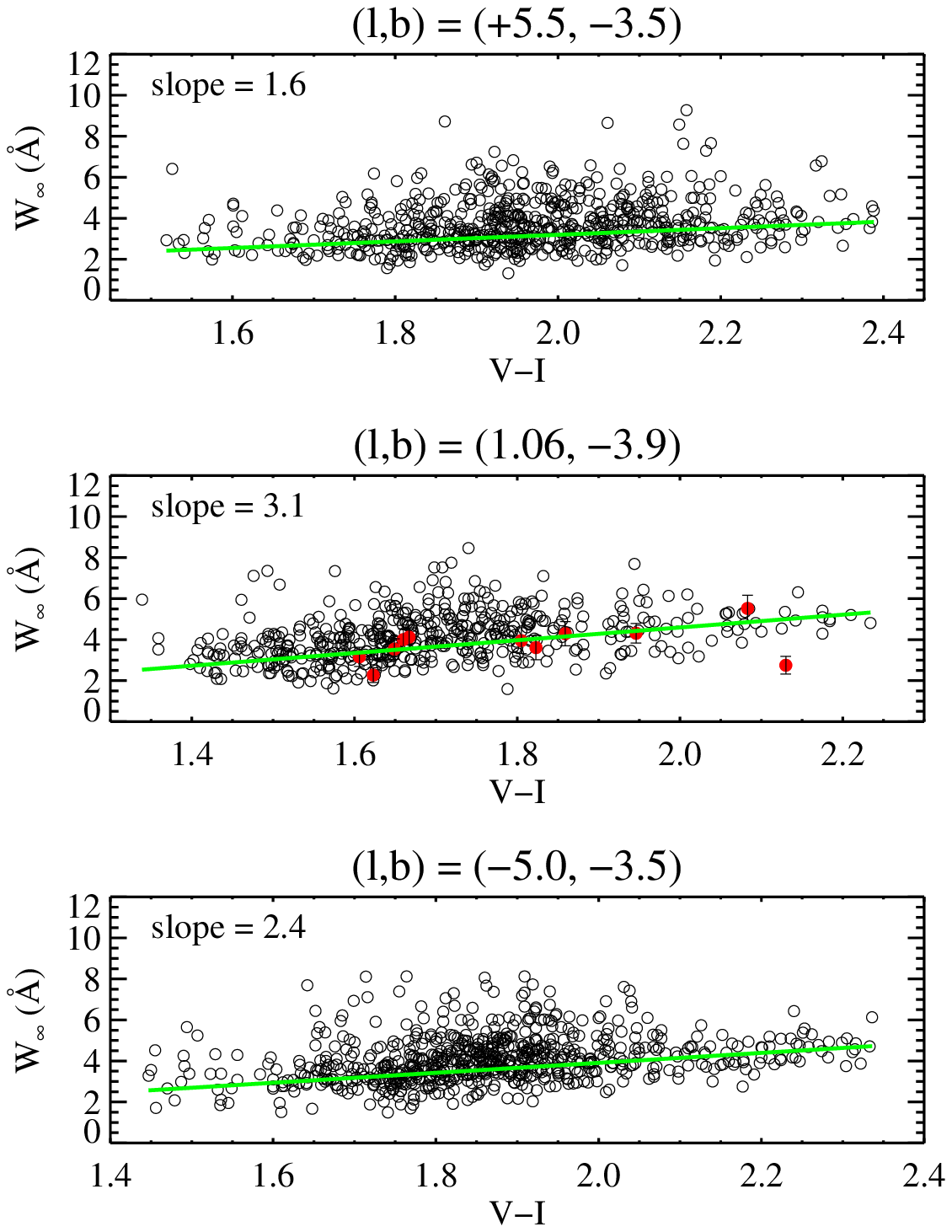}
  \caption{\voi\ as a function of color and I-band magnitude. The solid line is the best
  straight-line fit. Calibration stars indicated in red. \label{ewvi}}
\end{figure}

\begin{figure}
\center
    \includegraphics[scale=0.52]{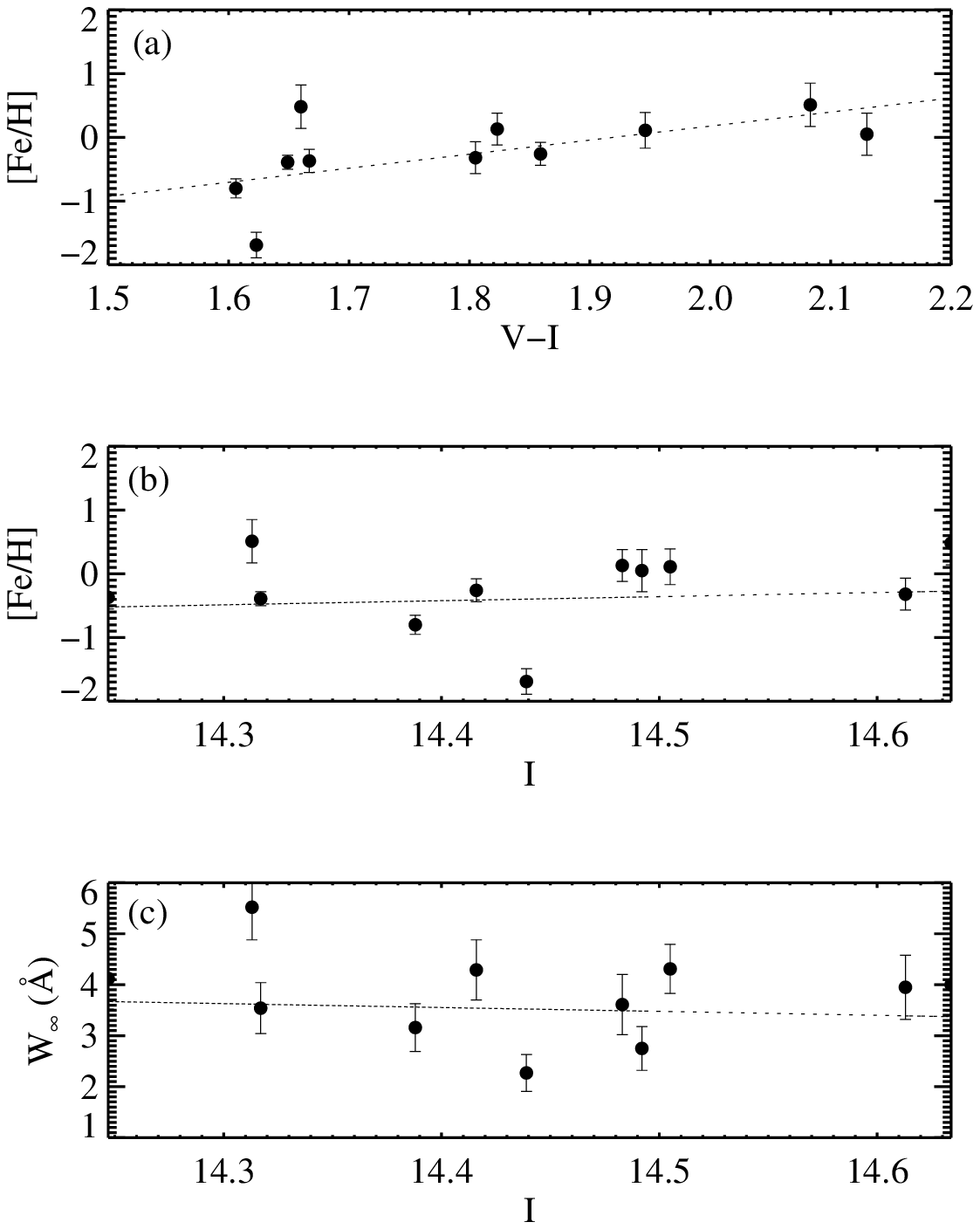}
    \includegraphics[scale=0.65]{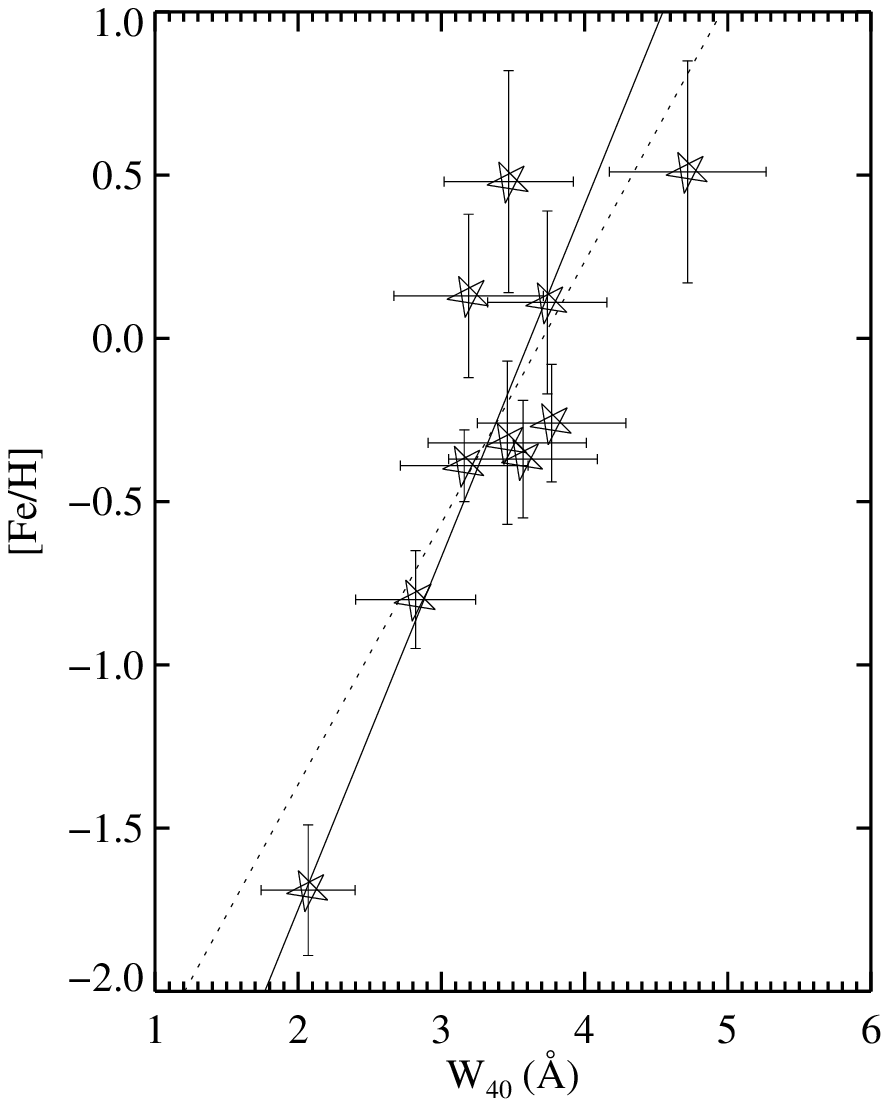}
    \caption{Left panel: (a) [Fe/H] as a function of V-I color; (b) I magnitude ; (c) EW as a function of
     I-band magnitude for the 10 calibration stars.
     Right panel: Calibration between [Fe/H] and CaT for the bulge RCGs. \label{fehca}}
\end{figure}

There are 11 bulge giants in our sample that have high-resolution [Fe/H] measurements from Zoc08.
We can use this sample to define a calibration between CaT and [Fe/H] for the bulge giants, and
then use this calibration to infer [Fe/H] for the rest of our sample. We identify these stars as
red clump giants, the metal-rich equivalent of horizontal branch stars, with a very narrow
intrinsic luminosity distribution. We do not need to correct to ``reduced equivalent width'' as we
did for the cluster giants in the previous section. To support this assumption we plot in Figure
\ref{ewvi} \voi\ as a function of I-band magnitude (left panels) for our full RCG sample at all
three LOS in the bar. There is no significant trend with luminosity. However, EW clearly does vary
with the V-I color as shown in right panels of Figure \ref{ewvi}. The slope of this relationship
changes with the position along the bar, and is largest in BW. The 11 common stars are shown in
red in Figure \ref{ewvi}, and exhibit a similar trend in [Fe/H], as shown in Figure
\ref{fehca}(a). Thus the change in EW with color in the RCGs reflects a real trend in metallicity,
and not a luminosity-dependent variation as seen on the red giant branch. Therefore we do not make
any corrections in our sample for this trend.

One of the common stars (with OGLEID 423286) has an anomalously low EW for a given (V-I) color
compared to the rest of the sample and will not be used in determining the calibration relation.
We plot [Fe/H] from Zoc08 against our $\mathrm{W}_{40}$ for the 10 remaining stars in the right
panel of Figure \ref{fehca}. The solid line shows the best straight-line fit to the data using the
uncertainties in both coordinates.

The calibrations for \car\ and \voi\ are:
\begin{equation}\label{cal2}
\mathrm{[Fe/H]} = (-3.828 \pm 0.131) + (0.475 \pm 0.154)(2.21\,\mathrm{W}_{40})
\end{equation}
and
\begin{equation}\label{cal3}
\mathrm{[Fe/H]} = (-3.545 \pm 0.125) + (0.384 \pm 0.114)(2.21 \mathrm{W}_{\infty})
\end{equation}
Here we separate the factor 2.21 for converting single-line EW measurement to the summed EW of the
triplet, to facilitate comparison with previous calibrations, e.g. Equation 2 above from C04.

This is the first determination of the CaT to [Fe/H] calibration for the bulge RCGs. All previous
measurements were done using red giants in globular clusters. The dotted line in the right panel
of Figure \ref{fehca} shows the globular cluster calibration (Equation \ref{cal}), which deviates
from the bulge calibration at low and high metallicities. The reduced $\chi^{2}$ for the globular
cluster calibration is greater by 1.5, suggesting that the bulge calibration may not the same as
the globular cluster calibration.

\citet{fulbright07} show that the [Ca/Fe] ratio in bulge red giants
varies from $\sim 0.4$ dex to 0.0 dex as metallicity increases from
[Fe/H] $\sim -1.5$ to $+0.0$ dex. Since our calibration is based on
the bulge RCGs, it will automatically compensate for this effect. If
the bulge and bar populations have different $\alpha$-element
enhancements, and the ratio of these populations vary with positon
along the bar, then our calibration, derived in BW, could
underestimate the metallicity at the $l = \pm
5$\degr\ positions. \citet{fulbright07} show that the difference in
$\alpha$-element enhancement between the bulge and the thick disk
populations is about 0.2 dex, so the ampliude of this underestimate
must certainly be less than this amount. Until high-resolution spectra
are obtained for stellar samples at positions other than BW, it will
be impossible to determine the exact change, if any, in the
calibration relation along the bar. It is highly unlikely that the
calibration changes by as much as 0.1 dex, an amount smaller than the
uncertainty of the current calibration.
\section{Stellar Metallicity Distributions}
\begin{figure}
\center
    \includegraphics[scale=0.65,angle=90]{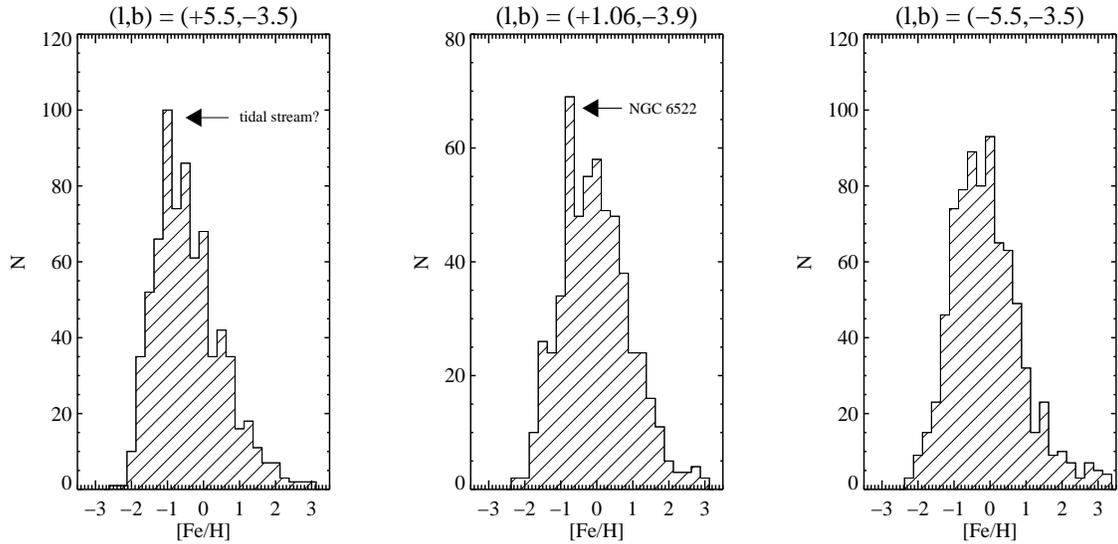}
 \caption{SMDs along the major axis of the bar.}\label{smd}
\end{figure}

We use \voi\ in Equation \ref{cal3} to infer metallicities for the RCG sample (all the points shown
in red in Figure \ref{cm}) to obtain the SMDs at three LOS in the bar. The SMDs are shown in
Figure \ref{smd}. In the BW SMD (middle panel) a distinct peak around $-0.90$ dex is from NGC 6522.
The distribution is significantly broadened by our relatively large uncertainties. To take into
account the effect of the individual errors we use maximum likelihood estimator \citep{pryor93} to
measure the true dispersion of these distributions. The mean and dispersion are listed in Table
\ref{ew} along with the corresponding kinematics as measured in Paper I. Only statistical
uncertainties are listed in this table. The calibration will introduce an additional systematic
uncertainty of about 0.10 dex.

Previous observations of \citet{fulbright06} \citep[also
  see][]{rich88,sadler96} show that the bulge metallicity distribution
can range from $-1.5$ to $+1.0$ dex. Our SMDs have extended high
metallicity tails that include $\sim 12$\% of the stars for each LOS
. Most of the stars in these tails are at the faint end of our sample
and have large EW uncertainties. Our calibrations extend only to
$\mathrm{[Fe/H]} = +0.5$ dex, and the analysis of section 3 suggests
that at higher metallicity the calibration may become non-linear.
Thus we do not believe that our observations give reliable evidence
for any stars with [Fe/H] $\gtrsim 1.0$ dex. This conclusion is
further supported by the comparison shown in Section 5.1 below, where
SMDs with maximum [Fe/H] $\lesssim$ 1.0 dex, when convolved with our
uncertainties, show similar extended tails.

\begin{figure}
    \center
    \includegraphics[scale=0.6,angle=90]{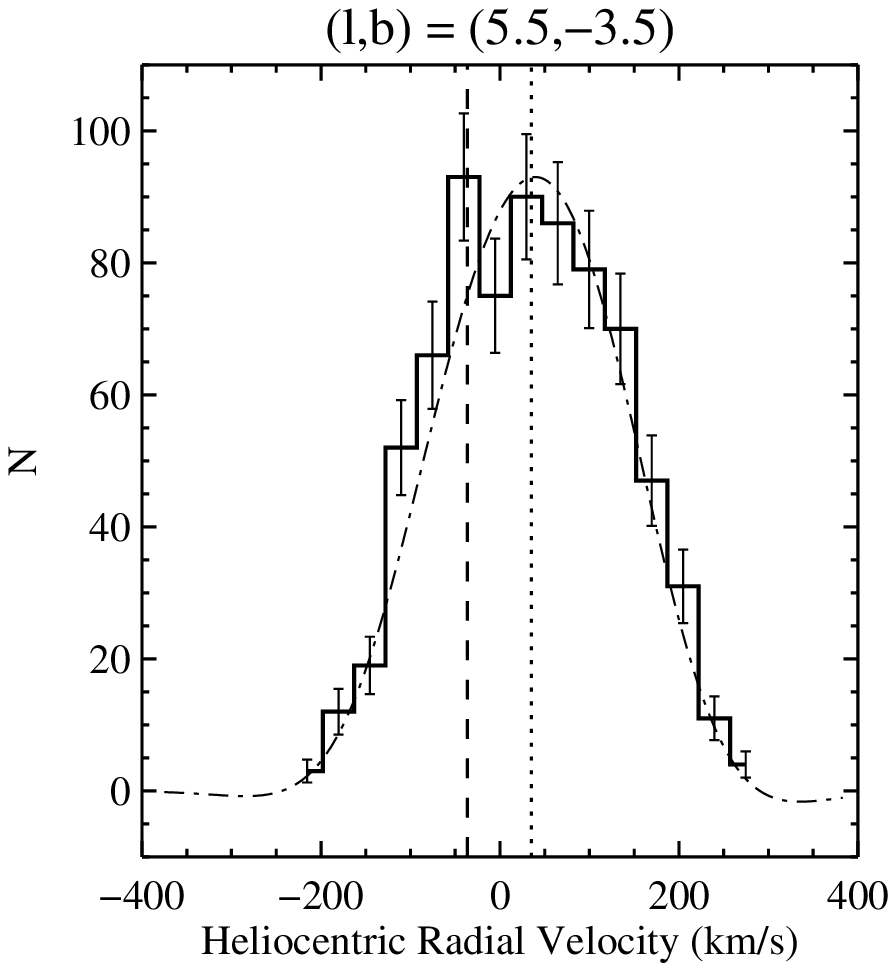}\\
    \includegraphics[scale=0.6]{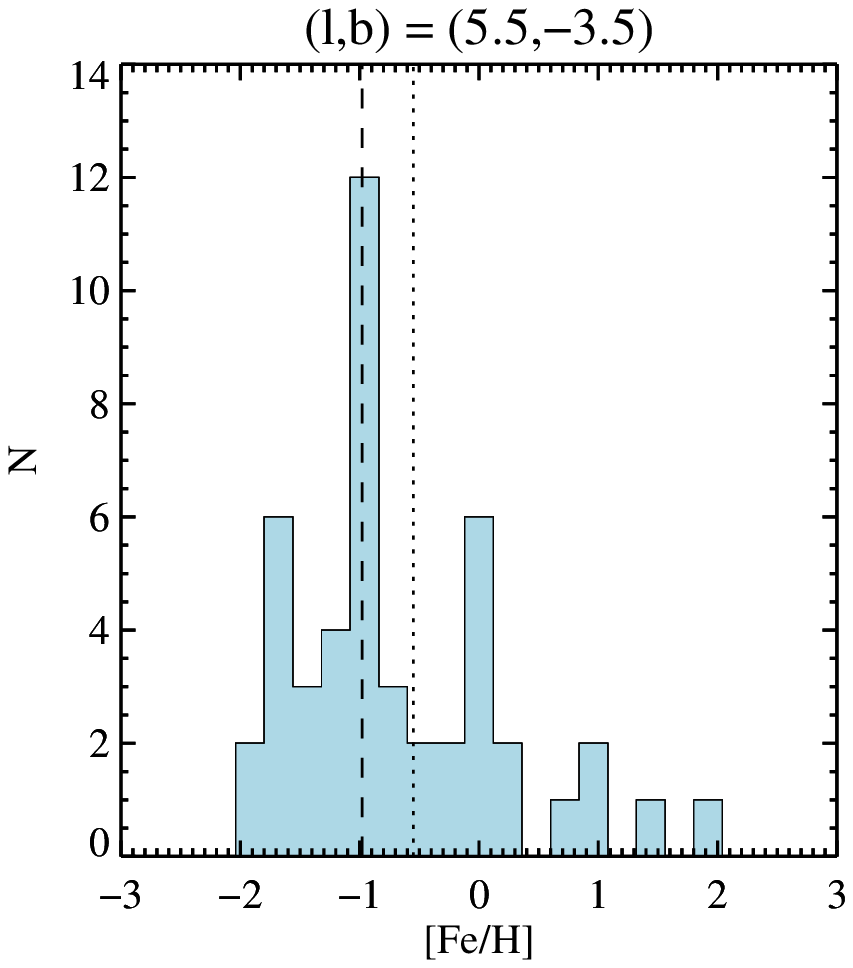}
    \caption{Possible signature of disrupted satellite: top panel: the velocity distribution of 738 stars, 
    with an excess around $-35$ \kms;
    bottom panel: the metallicity distribution of 47 stars in this velocity excess, selected with radial velocity between -45 \kms\ and  -28 \kms.
    In both panels the dotted line indicates the mean of the distribution and the dashed line indicates the location
    of the excess.\label{stream}}
\end{figure}

The SMD at $(l,b) = (5.5,-3.5)$ shows an excess at $\mathrm{[Fe/H]} =
-1.0$ dex, significantly different from the mean [Fe/H] of the
underlying distribution. The velocity distribution at this LOS from
Paper I, reproduced here in the top panel of Figure \ref{stream}, also
shows an excess at $\mathrm{V_{los}} = -36.5$ \kms, about 70
\kms\ away from the mean of the distribution. Since our measurements
of velocity are more accurate than those of metallicity, we select the
stars between $-45$ \kms\ and $-28$ \kms, i.e. within one velocity
standard deviation of the peak of this excess. This selection produces
a sample of 47 stars. The metallicity distribution of these stars is
shown in the bottom panel of Figure \ref{stream}, where the peak at
$-1.0$ dex is much more evident.  Other velocity selection limits
produce similar metallicity distributions, with the peak becoming less
striking as wider velocity limits are chosen. This feature may
indicate the presence of a tidal stream or debris from a disrupted
satellite.

Does this feature belong to the Sagittarius stream? The location and kinematics of the tidal
debris of the Sagittarius dwarf galaxy have been traced by \citet{majewski04} and
\citet{martinez04}. The Sagittarius stream passes through this LOS ($l,b = 5.5,-3.5$\degr), but is
located at heliocentric distance of $\sim 28$ kpc, on the other side of the Galaxy. The
metallicity of -1.0 dex of our feature is consistent with the range of metallicity of the stars in
the Sagittarius stream. However, the expected heliocentric radial
velocities of stars in the Sagittarius stream along our LOS is $+180 \pm 15$ \kms; the velocities
of the stars in our feature are $-36.5 \pm 8$ \kms. Thus we conclude that the cold kinematic
feature that we detect is not associated with the Sagittarius stream. To confirm the existence of
this stream and to better measure its properties will require additional spectroscopic
observations over a larger field of view.
\subsection{Comparison in Baade's Window}
\begin{figure}
    \center
 \includegraphics[scale=0.55]{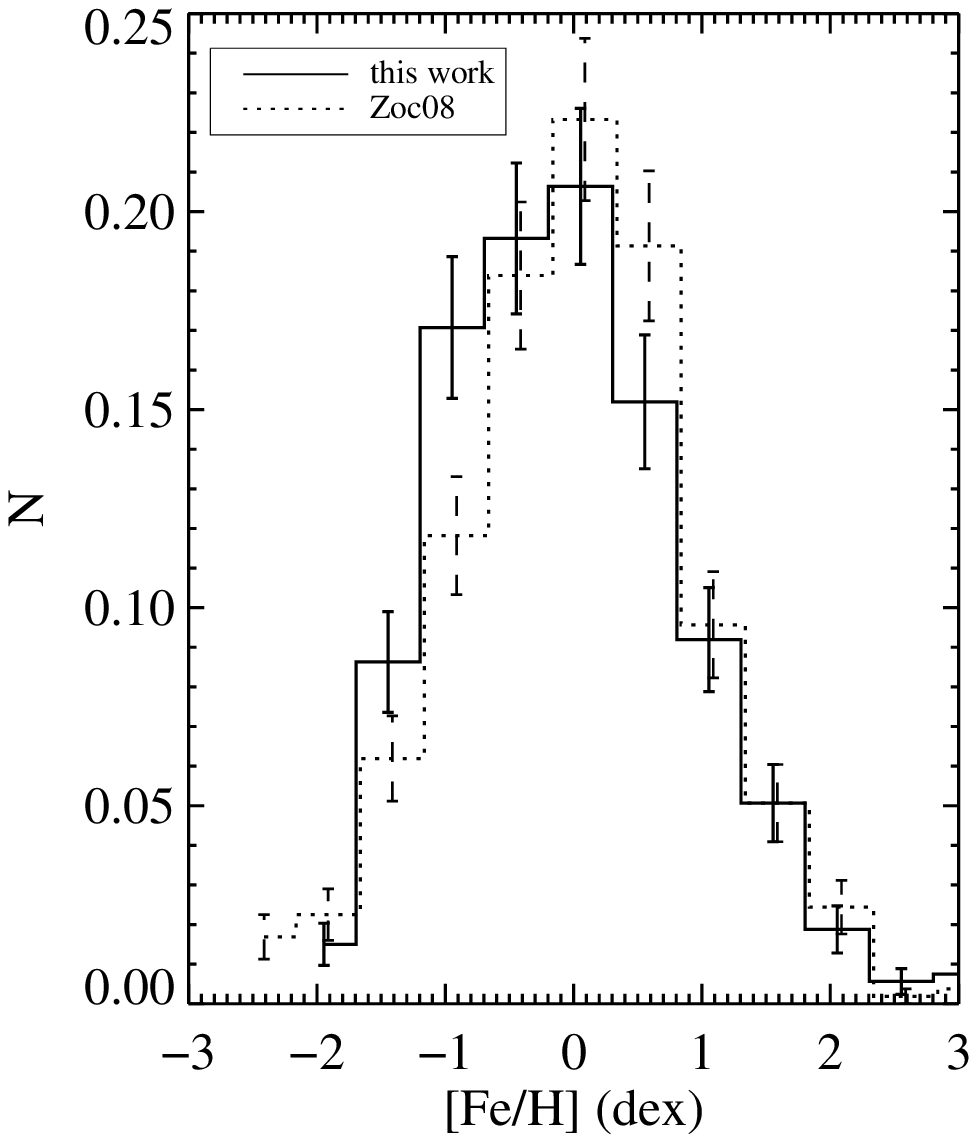} 
 \includegraphics[scale=0.55]{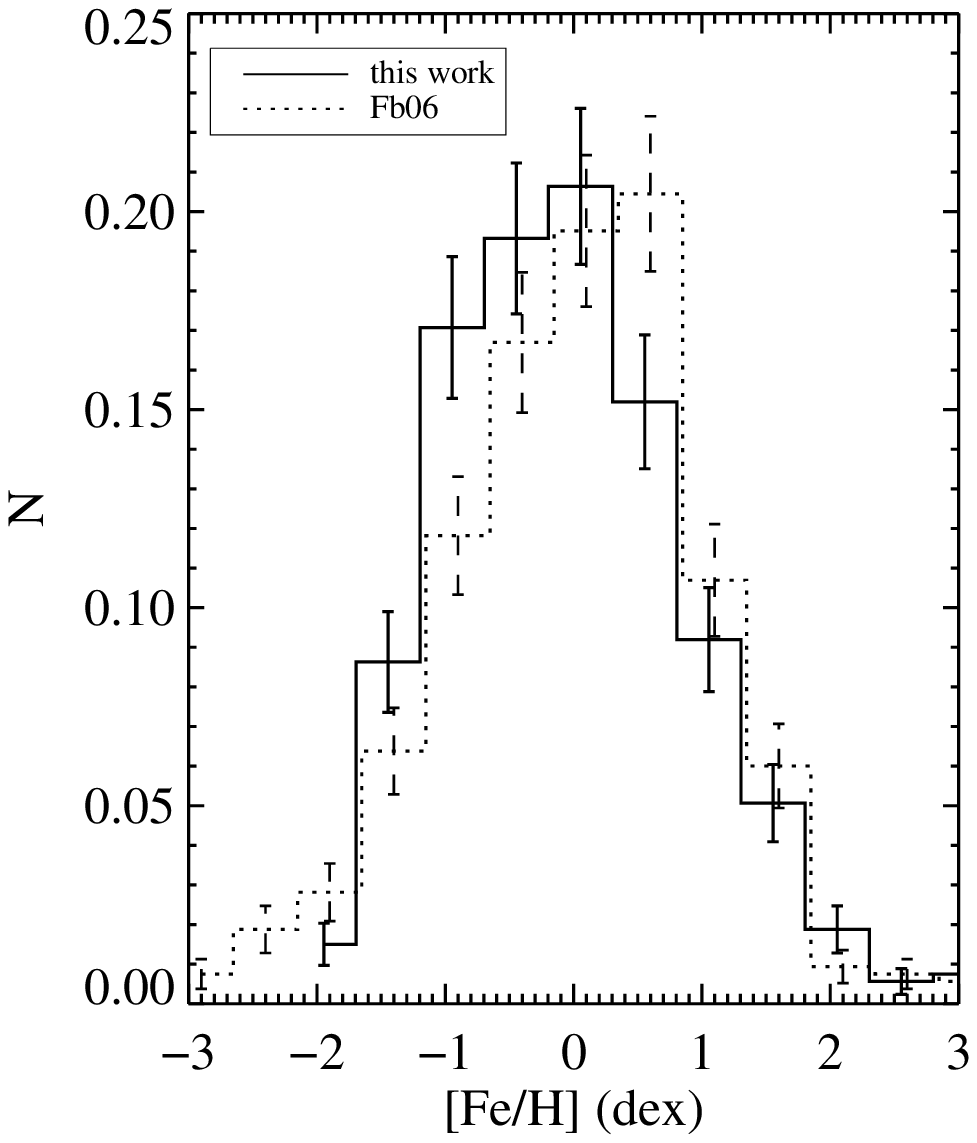} 
 \caption{Comparing the stellar metallicity distributions to previous work in Baade's Window.}\label{comp}
\end{figure}

We compare our SMD in BW with Zoc08 and \citet{fulbright06} (Fb06 hereafter) in Figure \ref{comp}.
Our BW/bulge sample consists of 557 RCGs. The Zoc08 sample consists of 204 stars at the bright end
of our RCG sample, and covers a very similar color range; their selection limits are indicated by
the rectangle in Figure \ref{cm}. They supplement their sample with 200 RCGs in BW  (Lecureur et
al. in prep.), but these latter measurements are not yet published so we can only make comparison
to the former sample. The Fb06 bulge sample is obtained by the recalibration of the
\citet{sadler96} sample that consists of about 320 RCGs.

Because the individual uncertainties of our sample are large, we convolve Zoc08 and Fb06 SMDs with
our error distribution to make consistent comparisons. We generate 10000 random realizations (with
replacement) of the Zoc08 and Fb06 samples, and add to them an error distribution that is randomly
drawn from our observed error distribution. The solid histogram in both panels of Figure
\ref{comp} is our SMD, and the dashed histograms show the Zoc08 (left panel) and Fb06 (right
panel) SMDs. We agree very well with both of these measurements. The major difference is an excess
of stars around $-0.90$ dex in our SMD, due to the cluster NGC 6522. Since the center of this
cluster is only 30 $\arcsec$ from the edge of our field, we include a significant number of
cluster stars in our sample. The sample of Fb06 discriminates against cluster stars by
construction \citep{terndrup95} and the photometric limits of Zoc08 select only a small part of
the cluster's red giant branch. Thus we expect our sample to include higher proportion of stars
from this cluster. Our mean metallicity in BW is $-0.09$ dex, close to solar, and agrees well with
Zoc08 ($\sim 0.0$ dex) and Fb06 ($-0.10$ dex). Our dispersion in BW is 0.58 dex, comparable to the
value of 0.52 dex calculated from the Fb06 sample. Both are larger than the dispersion of 0.40 dex
measured by Zoc08.
\subsection{Metallicity gradient along the Bar}
The mean metallicity and dispersion are lower at the two offset
positions in the bar compared to BW (see Table \ref{ew}). Zoc08 and
\citet{minniti95} find a gradient of about -0.25 dex along the minor
axis of the bar/bulge between b = $-4$\degr\ and $-12$\degr. We report
here the first measurements of metallicity gradients along the major
axis of the bar. We find a larger gradient towards $l = +5$
\degr\ ($-0.45 \pm 0.05$ dex) than towards $l = -5$\degr\ ($-0.10 \pm
0.05$ dex). The metallicity dispersion is larger in BW as compared to
the offset positions and decreases almost equally, by $-0.20 \pm 0.05$
dex in each direction. This trend correlates with the LOS velocity
dispersion that is also higher in BW and decreases by equal amounts at
the offset positions as shown in Paper I. The higher mean metallicity
at the center of bar may suggest that we are seeing an older, metal
rich bulge population in BW compared to the stars at $l = \pm
5$\degr\ in the bar.
\section{Discussion and Conclusion}
In this paper, we have presented EW measurements of the \ca\ line using FP absorption spectroscopy
at three position along the major axis of the Galactic bar. This is the first attempt to use FP
spectra for measuring stellar absorption line strengths. We show that reliable line strengths can
be obtained for hundreds of stars with this technique, allowing future FP observations to measure
accurate radial velocities and line strengths simultaneously from the same datacube.

We compare our CaT index with those of C04 and Ca07, and find that the major difference between
them comes from the differences in the integration bandpass. This has a minor effect for weak
lines but deviates non-linearly as the line strength increases, reflecting the strong damped wings
of the CaT lines. We find the three indices to be tightly correlated, and the ratio of EW to its
uncertainty remains constant for all of them. The differences between them does not affect any
previous metallicity measurement, since each index was calibrated to [Fe/H] appropriately. We have
investigated the effect of adding more samples to the FP data, and have found that the uncertainty
of the EW can be reduced by a factor of two (i.e. to be comparable to the accuracy of similar S/N
slit spectroscopy) by including as few as four extra samples on the red side of the absorption
line (between 8555 -- 8580 \AA).

To test the reliability our EWs we derive the metallicity of NGC 6522, a bulge globular cluster
that is present in our BW field. We used the C04 calibration to infer the metallicity from our
EWs. We obtained mean [Fe/H] for this cluster to be $-0.90$ dex, in excellent agreement with
recent measurement from Zoc08. However, previous measurements of NGC 6522 have reported a
relatively lower mean [Fe/H] of $-1.28 \pm 0.12$ \citep{terndrup98}, $-1.44 \pm 0.15$
\citep{zinn85} and $-1.21 \pm 0.04$ \citep{rutledge97}, where the former two measurements were
obtained using CMDs and the latter one used the CaT method. A relatively moderate mean
metallicity is surprising for this cluster as it has a very blue horizontal branch. This cluster
is most probably one of the `second parameter' bulge clusters, where the metallicity is not the
sole factor in determining the HB morphology \citep{catelan00, fusi97}. The second parameter could
be the age of the cluster, mass lost on the RGB, rotation, helium abundance, etc. The bulge sample
of Zoc08 also contains five stars from another bulge globular cluster NGC 6558 for which a detailed
analysis is presented in \citet{barbuy07}. This bulge cluster was also found to be moderately
metal poor ($-0.97$ dex) with a very blue horizontal branch. Their analysis suggested that age was
the second parameter for this cluster. They found it to be old at 14 Gyrs, and $\alpha$-enhanced.
A complete chemical abundance analysis of NGC 6522 is being performed by the same group (Barbuy et
al. in preparation). Any further analysis of second parameter issues in NGC 6522 is beyond the
scope of this paper.

We use the EWs of 10 stars in our sample that also have high
resolution [Fe/H] measurements to obtain the calibration between the
EW of the \ca\ line and [Fe/H] for the RCGs in the Galactic bulge. We
find this calibration to be marginally different from that of the
globular clusters. We use this calibration to infer [Fe/H] for all the
RCGs in our sample at three LOS in the bar to obtain the SMDs. The
signature of NGC 6522 is detected in the BW SMD around $-0.90$ dex. We
find the mean metallicity in BW to be close to solar, consistent with
previous work, and find a metallicity gradient that is different at
positive ($-0.45$ dex) and negative ($-0.10$ dex) longitudes. The metallicity
dispersion also shows gradient of about $-0.20$ dex between BW and $l =
\pm 5$\degr\ LOS. \citet{perez09} found that the bars in external
galaxies can have positive, negative, or null metallicity
gradients. They argue that these measurements can constrain the models
of structure formation and star formation history for the inner parts
disk galaxies, and may also distinguish between a bar and a classical
bulge.  We expect that our measurements can similarly constrain models
of the MW bar/bulge.

One would naturally expect that the metallicity gradient along the bar
would be symmetric with respect to the center. We observe that the
metallicity gradient is significantly different on the two sides of
the bar. We do not know the origin of this difference. One possible
reason would arise from extinction in the bar. The point of closest
approach of the LOS to the Galactic center lies behind the bar in the
first quadrant, but before the bar in the fourth quadrant. The density
of the metal rich bulge population decreases rapidly with distance
from the Galactic center. Thus if there is a significant amount of
extinction within the bar, as is commonly seen in other galaxies, the
LOS in the fourth quadrant will be more bulge-dominated than the LOS
in the first quadrant. Thus while there would be a metallicity
gradient on both LOS, the gradient in the first quadrant would be
greated, as observed.

The velocity and metallicity distributions at $l = +5.5$\degr\ show a distinct feature at $-36.5$
\kms\ and $-1.0$ dex respectively. This is a previously unknown cold kinematic feature, not
associated with the Sagittarius stream. Spectroscopic observation over a larger FOV along this LOS
are required to confirm this feature. This will be possible with our planned future survey of the bar.

We plan to use the FP system on the 10-m class Southern African Large Telescope (SALT) to obtain a
much more extensive determination of the kinematics and metallicity of stars in the inner Galaxy.
SALT's much greater aperture and larger FOV will enable us to measure $\gtrsim 2000$ stars along a
single LOS in an hour of observing time. We will investigate at least 10 LOS along the major axis
of the bar, obtaining 15000 -- 20000 stellar spectra in total. By slightly changing the observing
strategy for scanning the absorption line we will obtain EW measurements of much higher precision
than those of this paper. In addition we will select fields in BW that already have
high-resolution [Fe/H] measurements to greatly improve the precision of the CaT to [Fe/H]
calibration. Such a large sample of radial velocity and metallicities along several positions in
the bar will be ideal to look for other cold kinematic features from disrupted satellites, gradients
in the shape of velocity and SMDs, and correlations between them, and search for hypervelocity
stars.

\acknowledgements We thank Andrew Cole for his advice on the Calcium Triplet method and providing
the 47 Tuc data, Michael Rich for his very useful comments and suggestions, Manuela Zoccali for
confirming our NGC 6522 result, and Aaron Grocholski and Marta Mottini for their advice. Tad Pryor
provided a wealth of suggestions and insights. Support was provided by Andrew Baker, Eric Gawiser
and Saurabh Jha, and the National Science Foundation through grant AST0507323.
\newpage
\begin{table*}
\caption{Mean metallicity and its dispersion, and kinematics of the RCGs in the bar for three
LOS.}\label{ew}
\begin{tabular}{cccrcc}
 \tableline\tableline 
 LOS & $<\mathrm{[Fe/H]}>$ &
 $\sigma_{\mathrm{[Fe/H]}}$\tablenotemark{a} & $<\mathrm{V_{los}}>$ &
 $\sigma_{\mathrm{V_{los}}}$\tablenotemark{a} & N\\ (l,b) & (dex) &
 (dex) & $\mathrm{km~s^{-1}}$ & $\mathrm{km~s^{-1}}$ & \\ \tableline
 (5.5,-3.5) & $-0.55 \pm 0.03$ & $0.42\pm 0.03$ & $ 30.73 \pm 3.89$ &
 $102.38 \pm 2.39$ & 738\\ (1.1,-3.9) & $-0.09 \pm 0.04$ & $0.58\pm
 0.04$ & $-1.01 \pm 5.42$ & $112.62 \pm 3.41$ & 557\\ (-5.0,-3.5)&
 $-0.17 \pm 0.03$ & $0.39\pm 0.03$ & $-52.56 \pm 3.85$ & $102.30 \pm
 2.50$ & 804\\ \tableline
\end{tabular}
\tablenotetext{a}{The dispersion is measured using Maximum Likelihood Estimator}
\end{table*}
\newpage
\newpage
\newpage
\newpage
\newpage
\newpage
\newpage
\newpage
\newpage
\newpage
\newpage
\clearpage

\end{document}